\documentclass{desyproc}

\newcommand{\niceMKbabar}{\mbox{\sl B\hspace{-0.4em} {\small\sl%
A}\hspace{-0.37em} \sl B\hspace{-0.4em} {\small\sl%
A\hspace{-0.02em}R}}}

\begin{document}
\title{B Physics (Experiment)}

\author{{\slshape Michal Kreps} \\[1ex]
KIT, Wolfgang-Gaede-Stra{\ss}e 1, 77131 Karlsruhe, Germany }

\contribID{xy}  
\confID{1964}
\desyproc{DESY-PROC-2010-01}
\acronym{PLHC2010}
\doi            

\maketitle

\begin{abstract}
In past few years the flavor physics made important
transition from the work on confirmation the standard model of
particle physics to the phase of search for effects of a new
physics beyond standard model. In this paper we review
current state of the physics of $b$-hadrons with emphasis on
results with a sensitivity to new physics.
\end{abstract}

\section{Introduction}

The start of the $b$-physics dates back to 
1964 when
decay of the long lived kaon to two pions and thus the $CP$
violation was observed \cite{Christenson:1964fg}. It didn't
took very long until a proposal for theoretical explanation of
$CP$ violation was made. In their famous work, Kobayashi
and Maskawa showed that with 4 quarks there is no reasonable
way to include the $CP$ violation \cite{Kobayashi:1973fv}.
Together with it they also proposed several models to
explain the $CP$ violation in kaon system, amongst which the 6 quark
model got favored over time.

The explanation of the $CP$ violation in the six quark model
of Kobayashi and Maskawa builds on the idea
of quark mixing introduced by Cabibbo. The quark mixing
introduces difference between eigenstates of the strong and weak
interaction. 
The $CP$ violation
requires a complex phase  in order to provide a difference
between process and its charge conjugate.
In the four quark model, the quark mixing is described
by $2\times2$ unitarity matrix. With only four quarks, states
can be always rotated in order to keep the mixing matrix real
and thus quark mixing cannot accommodate the $CP$ violation.
Other arguments, which we are not going to discuss here,
prevent also suitable inclusion of the $CP$ violation in other
parts of the theory. With extension to six quarks, the mixing
matrix becomes $3\times3$ unitarity matrix, called
Cabibbo-Kobayashi-Maskawa matrix, $V_{CKM}$. In this case there is
no possibility to rotate away all phases and one complex
phase always remains in the matrix. This complex phase of
$V_{CKM}$ provides the $CP$
violation in the standard model. The idea has two important
implications. First, in addition to three quarks known in
early 1970's and predicted charm quark it postulates
existence of other two quarks, called bottom and top.
Second, despite the tiny $CP$ violation in the kaon system,
proposed mechanism predicts large $CP$ violation in the $B$
system. It took almost three decades, but both predictions
were experimentally confirmed, first by discovering the bottom
quark in 1977 \cite{Herb:1977ek} followed by the top quark
discovery in 1995 \cite{Abachi:1994td,Abe:1995hr} and
finally by the measurement of large $CP$ violation in the $B^0$
system in 2001 \cite{Aubert:2001sp,Abe:2001xe}.

In order to test the Kobayashi-Maskawa mechanism of the $CP$
violation many measurements are performed. In those main
aim is to determine the $V_{CKM}$ with a highest possible
precision. Tests are often presented in a form of the so
called unitarity triangle. It follows from the unitarity
requirement of the $V_{CKM}$. The product of the two
columns of the matrix has to be zero in the standard model. As
elements of the matrix are complex numbers, this requirement
graphically represents triangle in the complex plane. In the last
decade the flavor physics moved towards search for
inconsistencies which would indicate presence of a new
physics. We omit the charm mixing and $CP$ violation and
prospects of starting experiments which are discussed
elsewhere in these proceedings. Here we concentrate on the big
picture with some emphasis on tensions in various
measurements performed by \niceMKbabar{}, Belle, CDF, CLEO-c and
D\O{} experiments. 

\section{Sides of the unitarity triangle}

Looking to the unitarity triangle there are two sets of
quantities one can determine, namely angles and sides. In
this section we will discuss the status of sides determinations.
The sides itself are determined by  the $V_{td}$, $V_{ub}$
and $V_{cb}$ elements of the $V_{CKM}$. To determine those
quantities, two principal measurements are used. First type 
is the measurement of the $B^0$ oscillation frequency which
determines the $V_{td}$. Second type is the measurement of branching
fraction of semileptonic $B$ decays, which can be translated
to the $V_{ub}$ or $V_{cb}$. As there are no recent results on
the $B$ mixing, we concentrate on semileptonic decays and
determination of the $V_{ub}$ and $V_{cb}$.

The determination of the $V_{ub}$ and $V_{cb}$ is based on
the $b\rightarrow ul\nu$ and $b\rightarrow cl\nu$ transitions.
Advantage of semileptonic transitions is in confinement of
the all soft QCD effects into single form factor. In general two
complementary approaches exists. The first one is inclusive
measurements, where one tries to measure the inclusive rate of
the $B\rightarrow X_{(c,u)}l\nu$ rate with $X_{(c,u)}$ denoting
any possible hadron containing charm or up quark . The second approach uses
exclusive measurements where one picks up a well defined
hadron like $D^{*}$ in the case of $V_{cb}$ measurement. The two
approaches are complementary with inclusive being
theoretically clean in a first order, while exclusive being
much cleaner for experiment, but more difficult for theory.
In addition, part of the good properties of the inclusive
approach on the theory side is destroyed by a necessity of
kinematic requirements on the experimental side.
As one needs good control over background in those
measurements, it is practically domain of B-factories
running with the $e^+e^-$ at the $\Upsilon(4S)$ resonance.
 
Coming to the current status, determinations of the $V_{cb}$ as
well as the $V_{ub}$ has some issues and inconsistencies
\cite{Barberio:2008fa}.
In the inclusive determination of the $V_{cb}$ the fit to all
information has consistently too small $\chi^2$. On the
other hand in the exclusive determination using $B\rightarrow
D^*l\nu$ decays, different measurements are not fully
consistent with $\chi^2/ndf=56.9/21$. This inconsistency is
due to the differences between Belle and \niceMKbabar{} results
rather than inconsistence between old and new measurements. 
The world average determined from the inclusive measurement is
$V_{cb}=(41.5\pm0.44 \pm0.58)\cdot 10^{-3}$, from the $B\rightarrow
Dl\nu$ we obtain $V_{cb}=(39.4\pm1.4\pm0.9)\cdot 10^{-3}$
and from the $B\rightarrow D^*l\nu$
$V_{cb}=(38.6\pm0.5\pm1.0)\cdot 10^{-3}$. As can be seen,
despite the tension in the experimental information from
$B\rightarrow D^*l\nu$ decays, the two exclusive determinations
agree with each other, but the inclusive approach yields value
which is about $2.3\sigma$ higher than the one from exclusive
determination.

While the determination of the $V_{ub}$ is in principle same as
the determination of the $V_{cb}$, in practice the $V_{ub}$ is much more
difficult due to the smallness of the $b\rightarrow ul\nu$ branching
fraction compared to the $b\rightarrow cl\nu$. The $b\rightarrow
cl\nu$ in this case is a significant background. Kinematic
selection to reduce this background destroys possibilities
of theory for the precise and reliable calculations. On
the inclusive determination side, there are several groups which perform fits
to the experimental data of inclusive decays. 
\begin{figure}
\begin{minipage}[t]{0.34\textwidth}
\centerline{\includegraphics[width=0.65\textwidth]{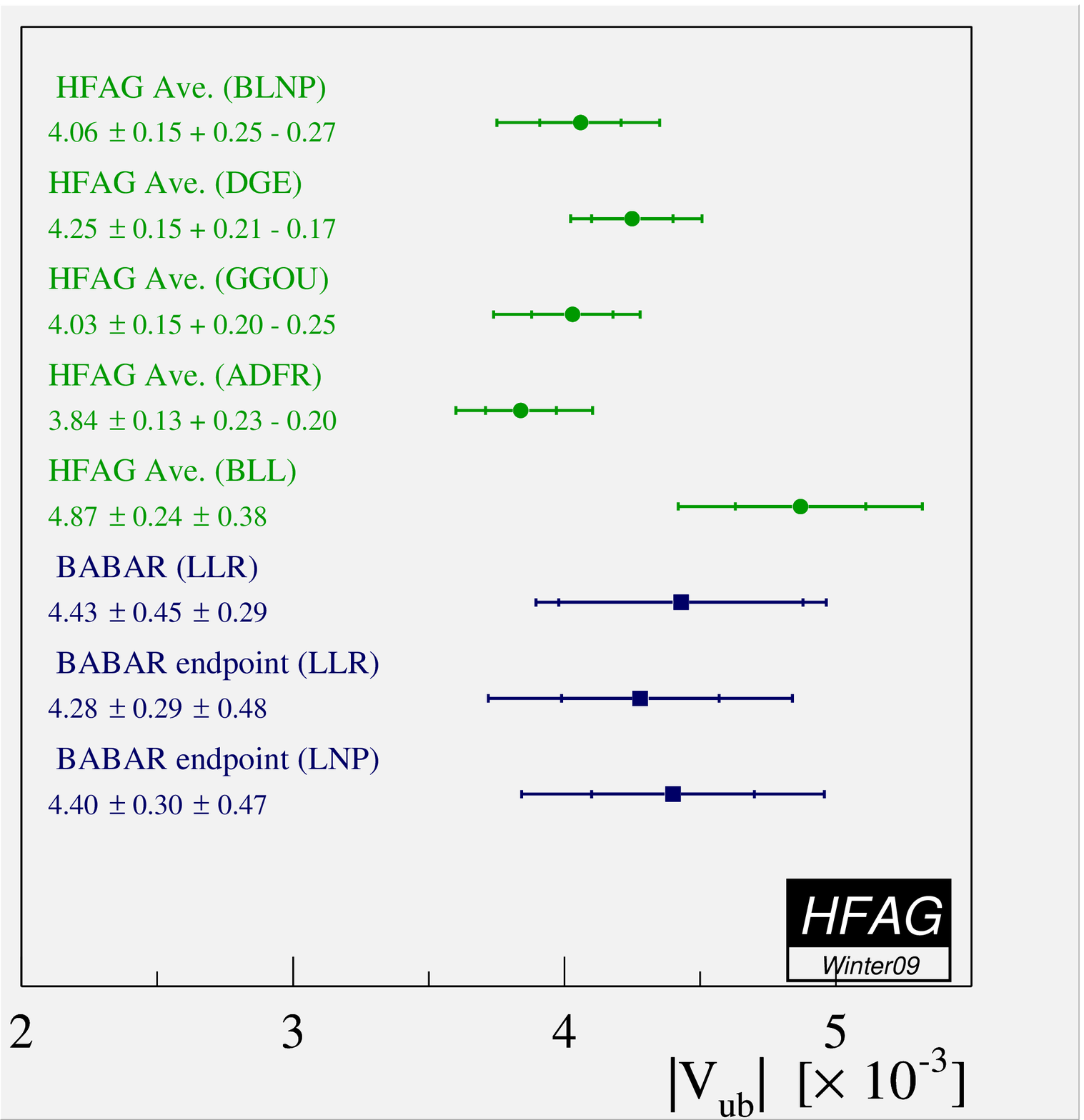}}
\caption{Summary of different inclusive determinations of
the $V_{ub}$ from semileptonic $b\rightarrow ul\nu$ decays
\cite{Barberio:2008fa}.}\label{fig:Vub}
\end{minipage}
\hfill
\begin{minipage}[t]{0.60\textwidth}
{\includegraphics[width=0.37\textwidth]{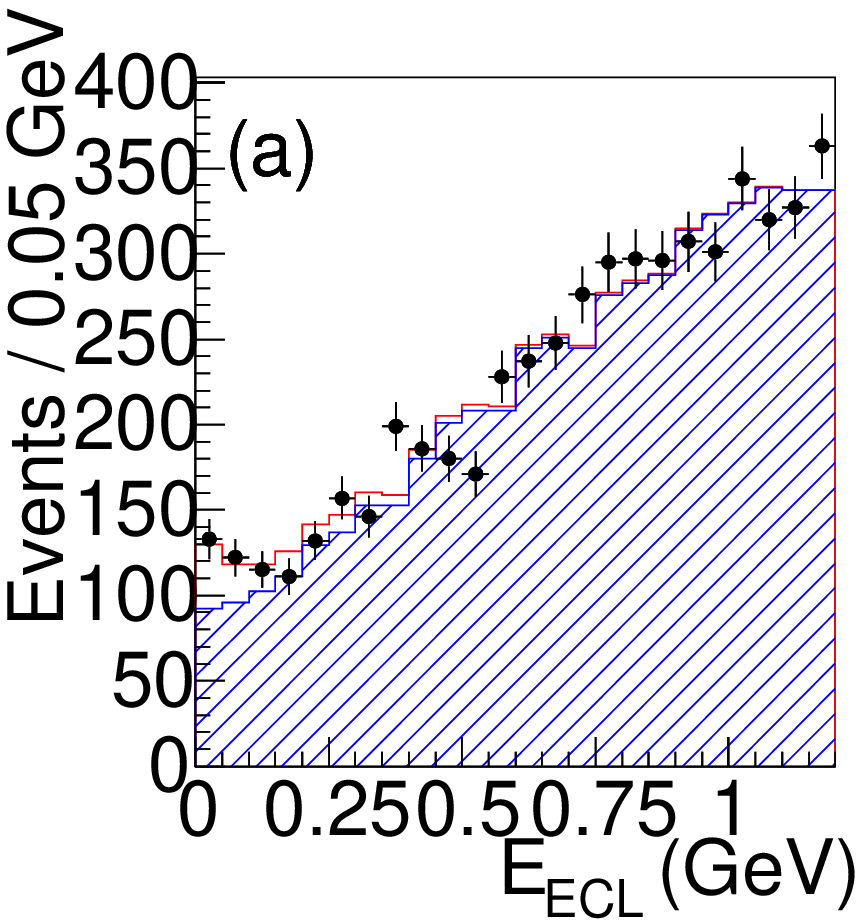}
          \hfill \includegraphics[width=0.60\textwidth]{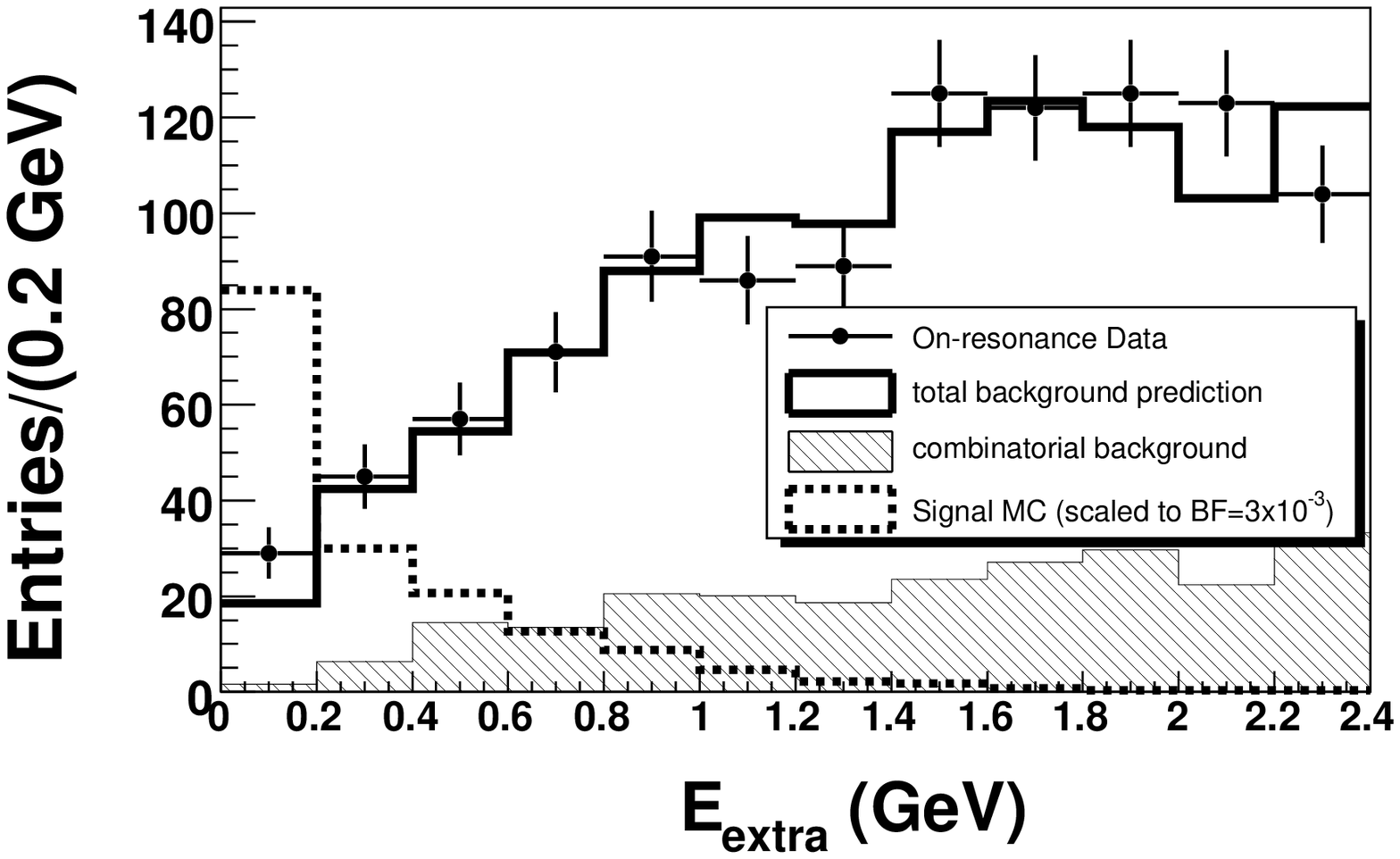}}
\caption{Distribution of the remaining energy in
the $B\rightarrow\tau\nu$ searches using semileptonic tag at
Belle (left) and fully hadronic tag at \niceMKbabar{} (right).}\label{fig:Btaunu}
\end{minipage}
\end{figure}
On the exclusive side, \niceMKbabar{} experiment provides new result on
the $B\rightarrow\pi l\nu$ and $B\rightarrow\rho l\nu$. Using
their partial branching fraction in different momentum
transfer regions together with lattice QCD calculations they
derive $|V_{ub}|=(2.95 \pm 0.31) \times 10^{-3}$
\cite{Aubert:2010uj}, which is about $2\sigma$ below
the inclusive determinations. If this stands, than we have
another discrepancy in the sides of unitarity triangle.

Another way of accessing $V_{ub}$ is to use
$B^+\rightarrow\tau\nu$ leptonic decays which proceed
through weak annihilation. In the standard model its rate is
given by expression
\begin{equation}
BF=\frac{G_F^2m_B}{8\pi}m_l^2\left(1-\frac{m_l^2}{m_B^2}\right)^2f_B^2|V_{ub}|^2\tau_B,
\end{equation}
where all quantities except of $f_B^2$ and $V_{ub}$ are well
known. Typically one takes input on the $f_B^2$ and $V_{ub}$
from other measurements and puts constraints on a new physics.
Alternatively one can take measured branching fraction
together with the prediction for $f_B^2$ and extract $V_{ub}$. 
B-factories provided recently evidence for this decay. Both,
Belle and \niceMKbabar{} reconstruct one $B$ in a semileptonic
or a fully
hadronic decay, called tagged, together with identified charged products of
the $\tau$ decay. In such events, all what should be remaining
are neutrinos and therefore one expects zero additional
energy in the event. In Fig.~\ref{fig:Btaunu} we show
examples of the distribution of additional energy. The Belle
experiment sees evidence on the level of $3.5\sigma$ in both
tags \cite{Ikado:2006un,Adachi:2008ch} while \niceMKbabar{} experiment obtains excess of about
$2.2\sigma$ \cite{Aubert:2007xj,Aubert:2008gx}.
The world average of the branching fraction of
$(1.73\pm0.35)\cdot 10^{-4}$ is little higher than the SM
prediction of $(1.20\pm0.25)\cdot 10^{-4}$ and yields
$V_{ub}$ which is in some tension with other determinations.

\begin{figure}
\begin{minipage}[t]{0.38\textwidth}
\centerline{\includegraphics[width=0.8\textwidth]{kreps_michal.fig3.eps}}
\caption{Comparison of the predicted $f_{D_s}$ with experimental
results. The circles denote experimental values with the yellow band
showing average. The squares show prediction and the gray area
theory average. The green lines denote the difference between theory
and experiment in Gaussian $\sigma$. The time $t$ is
measured since June 2005.}\label{fig:fds}
\end{minipage}
\hfill
\begin{minipage}[t]{0.60\textwidth}
\includegraphics[width=0.49\textwidth]{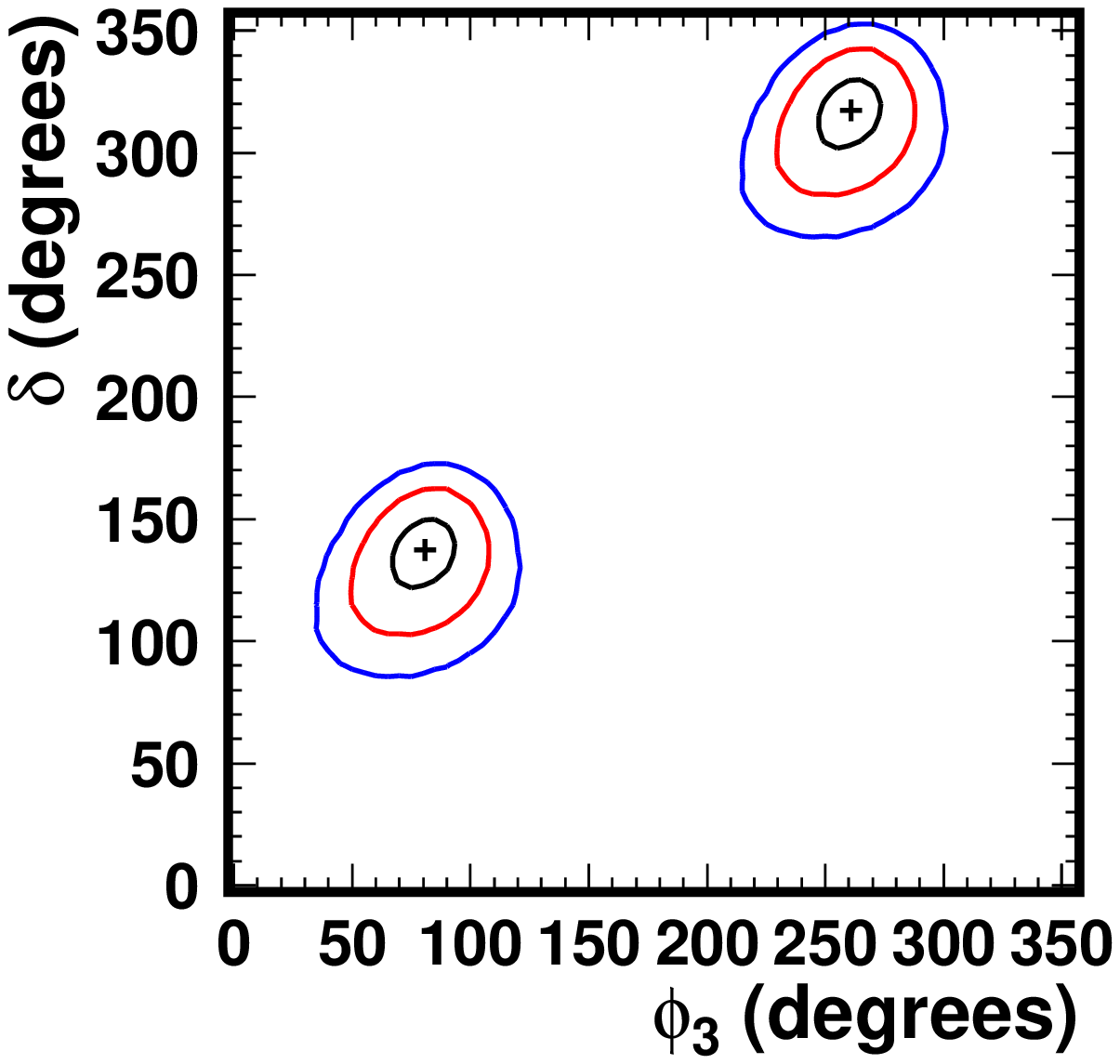}
\hfill
\includegraphics[width=0.49\textwidth]{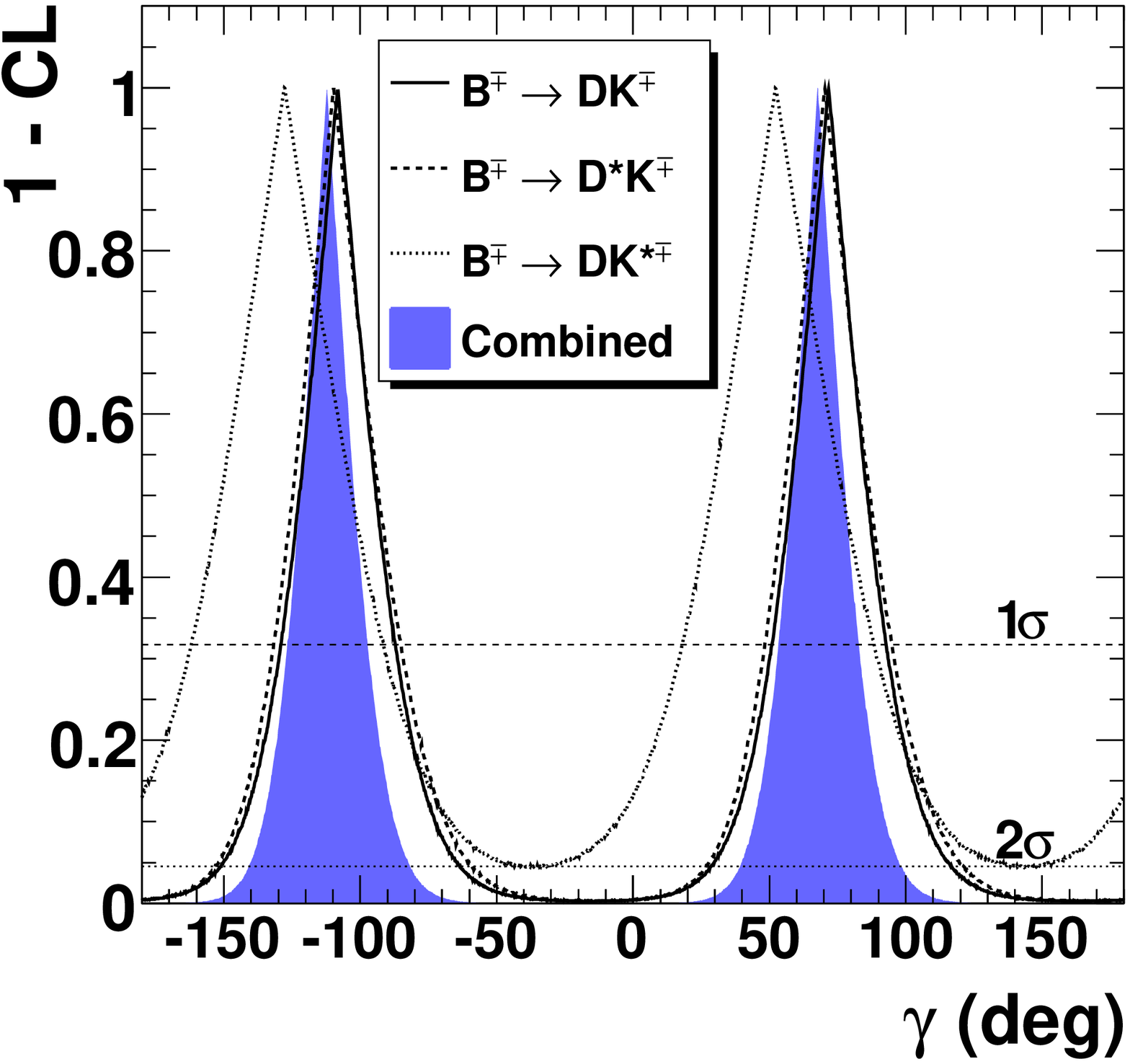}
\caption{Confidence regions in the plane of strong phase
$\delta$ and the CKM angle $\gamma/\phi_3$ from Belle experiment
(left) and 1-CL for the CKM angle $\gamma$ from \niceMKbabar{} experiment
(right). On the left plot, contours correspond to 1, 2 and 3
standard deviations. On the right plot, separate contours for
decays $B^+\rightarrow D^0K^+$, $B^+\rightarrow D^{*0}K^+$,
$B^+\rightarrow D^0K^{*+}$ and combination of all is shown.}\label{fig:gamma}
\end{minipage}
\end{figure}

The result of the $B^+\rightarrow\tau\nu$ branching fraction
brings up the question whether theory prediction from
the lattice QCD for the $f_B^2$ is correct. One way to test
predictions is to turn to charm sector where we expect
smaller contributions from a new physics. Decay
$D_s^+\rightarrow \tau^+\nu$ is a usual testing ground for
calculations.
The branching
fraction is given by same formula as for
$B^+\rightarrow\tau\nu$ with replacing $f_B^2$ and $V_{ub}$
by their appropriate counterparts. The branching fraction for $D_s^+\rightarrow
\tau^+\nu$ was measured by CLEO, \niceMKbabar{} and Belle experiments
and there used to be some discrepancy between the prediction for
$f_{D_s}$ and its value extracted from the $D_s^+\rightarrow
\tau^+\nu$ data. Summary of the evolution of this
discrepancy is shown in Fig.~\ref{fig:fds}
\cite{Kronfeld:2009cf}.
Current situation is not too critical anymore as
the discrepancy went down from $4\sigma$ to $2\sigma$.
With this we conclude discussion of sides of the unitarity
triangle, where despite lot of the experimental work and large
progress several tensions remains.

\section{Angles of the unitarity triangle}

The angles of the unitarity triangle are defined as
\begin{eqnarray}
\alpha&=&\arg\left(-{V_{td}V^*_{tb}}/{V_{ud}V^*_{ub}}\right),\\
\beta&=&\arg\left(-{V_{cd}V^*_{cb}}/{V_{td}V^*_{tb}}\right),\\
\gamma&=&\arg\left(-{V_{ud}V^*_{ub}}/{V_{cd}V^*_{cb}}\right).
\end{eqnarray}
As they are give by the phases of complex numbers,
their determination is possible only through $CP$ violation
measurements. Here we omit determination of the angle $\alpha$,
briefly mention status of the angle $\beta$ and concentrate on
the angle $\gamma$ which received most of the new experimental
information.

The angle $\beta$ is practically give by the $V_{td}$ phase. One
of the process where this CKM matrix element enters is the $B^0$
mixing. Its best determination comes from the measurement of
$CP$ violation due to the interference of decays with and
without mixing to a common final state. Using decays to
$c\overline c$ resonance with neutral kaon \niceMKbabar{} extracts using final dataset $\sin
2\beta=0.687\pm0.028\pm0.012$ \cite{Aubert:2009yr}. The latest
measurement from Belle experiment gives
$\sin2\beta=0.642\pm0.031\pm0.017$ \cite{Chen:2006nk}. It is worth to
note that both experiments are still statistically limited. 

Determination of the angle $\gamma$ provides important
information for tests of a physics beyond standard model. It
is determined from the interference of tree level
$b\rightarrow c$ and $b\rightarrow u$ transitions and thus
having small sensitivity to a new physics. While several
different decays are suggested for the determination, all current
experimental information comes from the $B^+\rightarrow D^0K^+$.
In those decays, the $b\rightarrow c$ transition provides
$B^+\rightarrow D^0K^+$ decay while the $b\rightarrow u$
transitions yields $B^+\rightarrow \overline{D}^0K^+$ final
state. Thus measurement of the $CP$ violation in the final
states which are common to $D^0$ and $\overline{D}^0$ is
needed. Three  different final states are currently used.
The first one uses q Cabibbo favored $\overline{D}^0\rightarrow
K^-\pi^-$ with q doubly Cabibbo suppressed $D^0\rightarrow
K^-\pi^+$ \cite{Atwood:1996ci,Atwood:2000ck}. The second method
uses a Cabibbo suppressed $D^0$ decays like $\pi^+\pi^-$,
$K^+K^-$ \cite{Gronau:1990ra}. The third approach uses a Dalitz
plot analysis of a $D^0\rightarrow K_s\pi^+\pi^-$
\cite{Giri:2003ty}.
The main limitation is that rates are small and up to now there
was no significant measurement of the $CP$ violation in those
decays. 
Recently Belle and \niceMKbabar{} experimental announced
$\approx 3.5\sigma$ evidence for the $CP$ violation in the
$B^+\rightarrow D^0K^+$ decays with $D^0\rightarrow
K_s\pi^+\pi^-$ \cite{delAmoSanchez:2010rq,Poluektov:2010wz}.
The extracted
confidence regions on the angle $\gamma$ are shown in
Fig.~\ref{fig:gamma}. Belle experiment extracts
$\gamma=(78^{+11}_{-12}\pm4\pm9)^\circ$ and \niceMKbabar{} obtains
$\gamma=(68\pm14\pm4\pm3)^\circ$. 

\section{$B_s$ sector}

The $CP$ violation in the $B_s$ meson sector is currently
the most exciting place and widely discussed in relation to
a new
physics. Two results, which are in many models of a new
physics related are the measurement of the $CP$ violation in the
$B_s\rightarrow J/\psi\phi$ decays and the flavor specific
asymmetry in a semileptonic $B_s$ decays. 

\begin{figure}
\centerline{\includegraphics[width=0.30\textwidth]{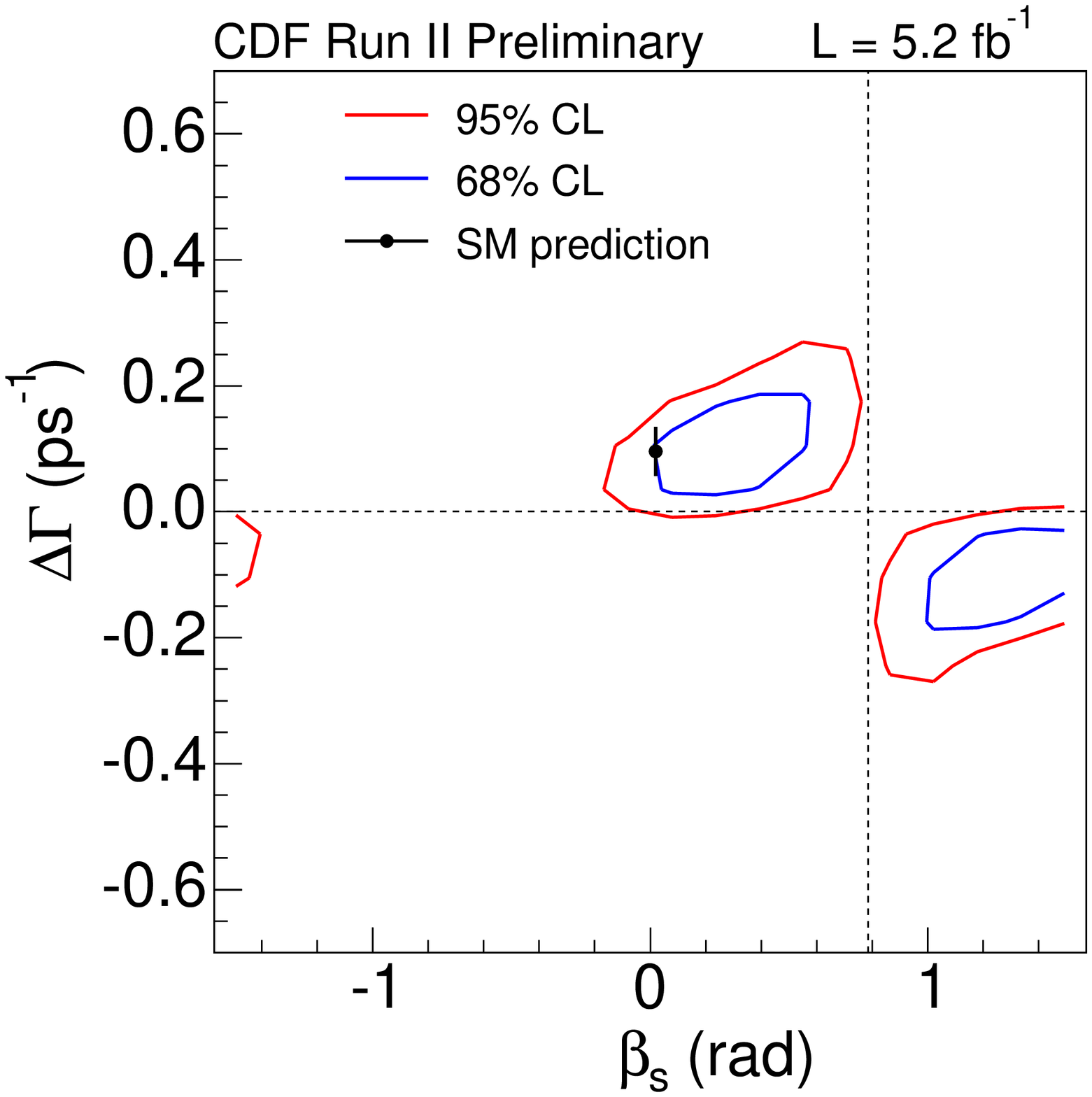}
  \hspace{0.1\textwidth}\includegraphics[width=0.30\textwidth]{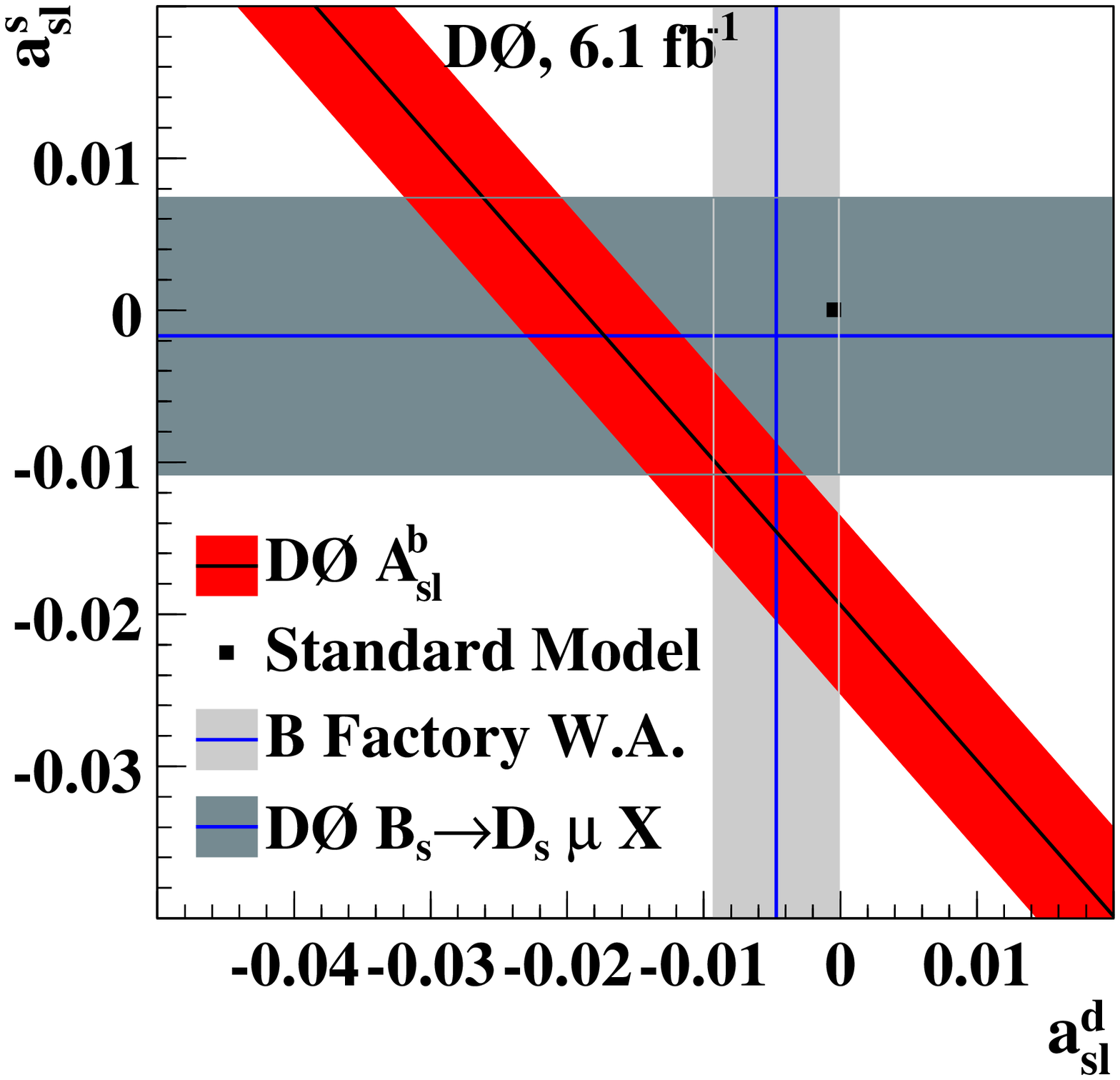}}
\caption{The $\Delta\Gamma_s$-$\beta_s$ confidence regions in
$B_s\rightarrow J/\psi\phi$ decays from CDF experiment using
$5.2$ fb$^{-1}$ of data (right). Latest results on the
flavor specific asymmetry in semileptonic $B_{(s)}$ decays
from D\O{} experiment (right).}\label{fig:bs}
\end{figure}

The origin of the
first one is in the interference of the decays with and
without $B_s$ mixing. The standard model predicts only tiny
$CP$ violation which comes from the fact that all CKM matrix
elements entering are almost real. The previous results from
Tevatron experiments showed about $1.5$-$1.8\sigma$
deviation from the standard model \cite{cdf:betas,d0:betas} with combination being
$2.2\sigma$ away. Recently CDF collaboration updated
its result with more data and few improvements, which yield
the better constraints on the $CP$ violation in
$B_s\rightarrow J/\psi\phi$. Resulting 2-dimensional
$\Delta\Gamma_s$-$\beta_s$ contour is shown in
Fig.~\ref{fig:bs}. Overall CDF experiment now observes better
agreement between the data and standard model with
difference of about
$0.8\sigma$. More details on this update can be find
in Ref.~\cite{proc:Elisa}.

The second measurement we present here is measurement of the
flavor specific asymmetry in semileptonic $b$-hadron decays.
In the standard model as well as in a large class of new
physics models this quantity is predicted to be small. It
can be generated or by a direct $CP$ violation or by an
asymmetry in the mixing rate between $b$- and
$\overline{b}$-mesons. Typically direct $CP$ violation is
zero as we talk about the most allowed decay amplitude
$b\rightarrow cl\nu$ which would need
a second contribution to interfere with. As it is
not easy to construct a model where a second amplitude with
reasonable size would exists typically the direct $CP$ violation
is predicted to be zero. The effect of different mixing
rates is small for the $B^0$ due to the small decay width difference
and small in the standard model for the $B_s$ due to the small
phase involved. D\O{} experiment announced a new measurement
this year, with a highly improved treatment of systematic
uncertainties. They measure $A_{fs}^b=(-96\pm25\pm15)\times
10^{-4}$ which is significantly different from the standard
model expectation of $A_{fs}^b=(-2.3^{+0.5}_{-0.6})\times
10^{-4}$ \cite{Lenz:2006hd}. If this result is confirmed, it is clear sign of
the physics beyond the standard model. For more details see Ref.~\cite{proc:bertram}. 

\section{Rare decays}

Rare FCNC transitions are best known outside the flavor physics
community for searches of a physics beyond standard model.
Prime example is rare $B_s\rightarrow\mu^+\mu^-$
decay, where previous results could put strong constraints
on some new physics model, even with limits, which are far
from the standard model expectations. The standard model prediction
for the branching fraction of $B_s\rightarrow\mu^+\mu^-$
is $(3.6\pm0.3)\times 10^{-9}$ \cite{Buras:2009us}.
The main difficulty is in suppressing and controlling
background. The search for this decays is dominated by the
Tevatron experiments. Recently D\O{} experiment updated
their result using 6.1 fb$^{-1}$ of data which yields upper
limit on the branching fraction of $5.2\cdot 10^{-8}$ at 95\%
C.L. \cite{Abazov:2010fs}. The best limit at  this moment is one from the CDF
experiment using 3.7  fb$^{-1}$ of data and the upper limit of
$4.3\cdot 10^{-8}$ at 95\% C.L. \cite{cdf:bsmumu}. Those are
about an order of magnitude above the standard model prediction.

\begin{figure}[th]
\centerline{\includegraphics[width=0.25\textwidth]{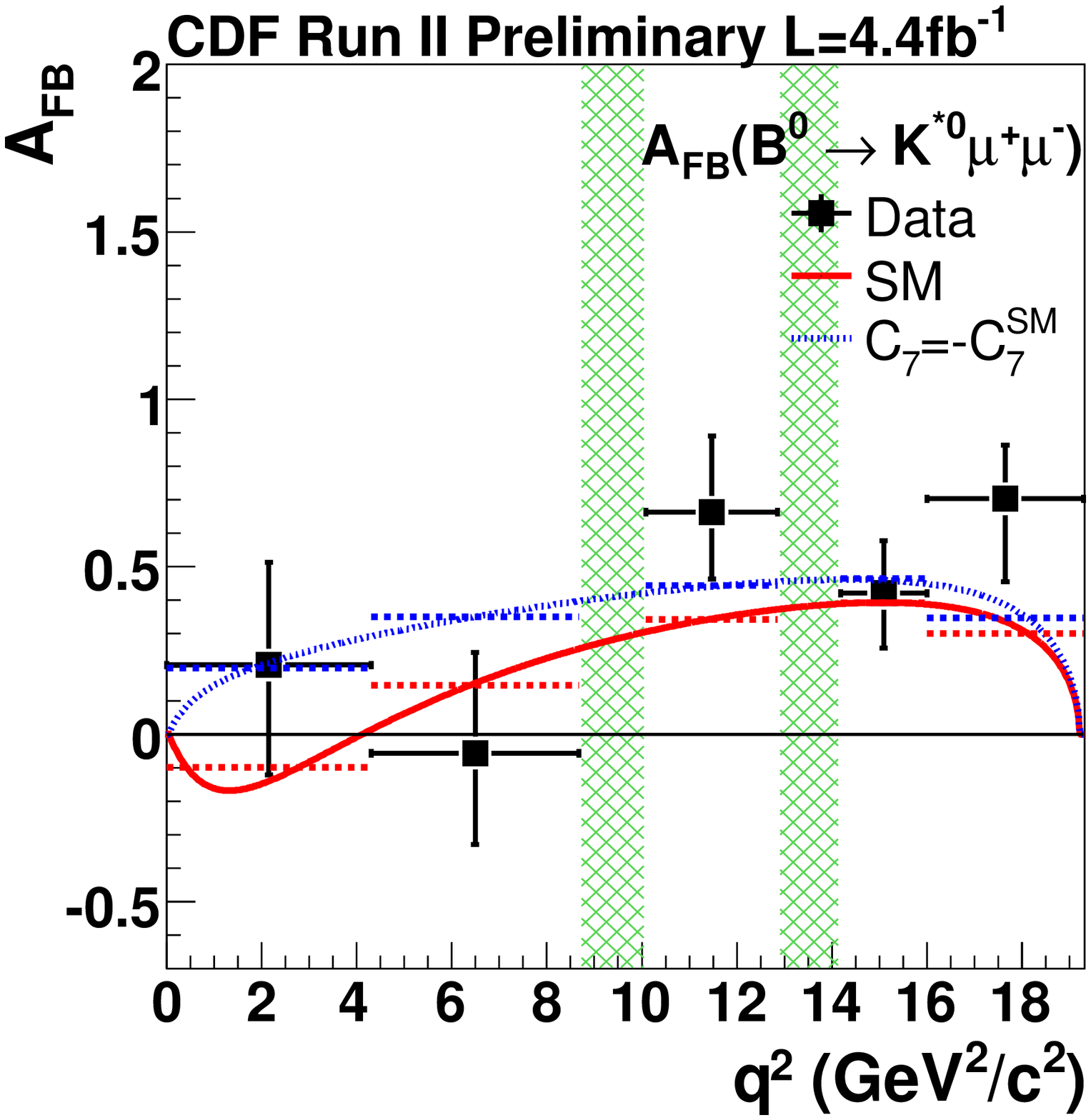}
  \includegraphics[width=0.30\textwidth]{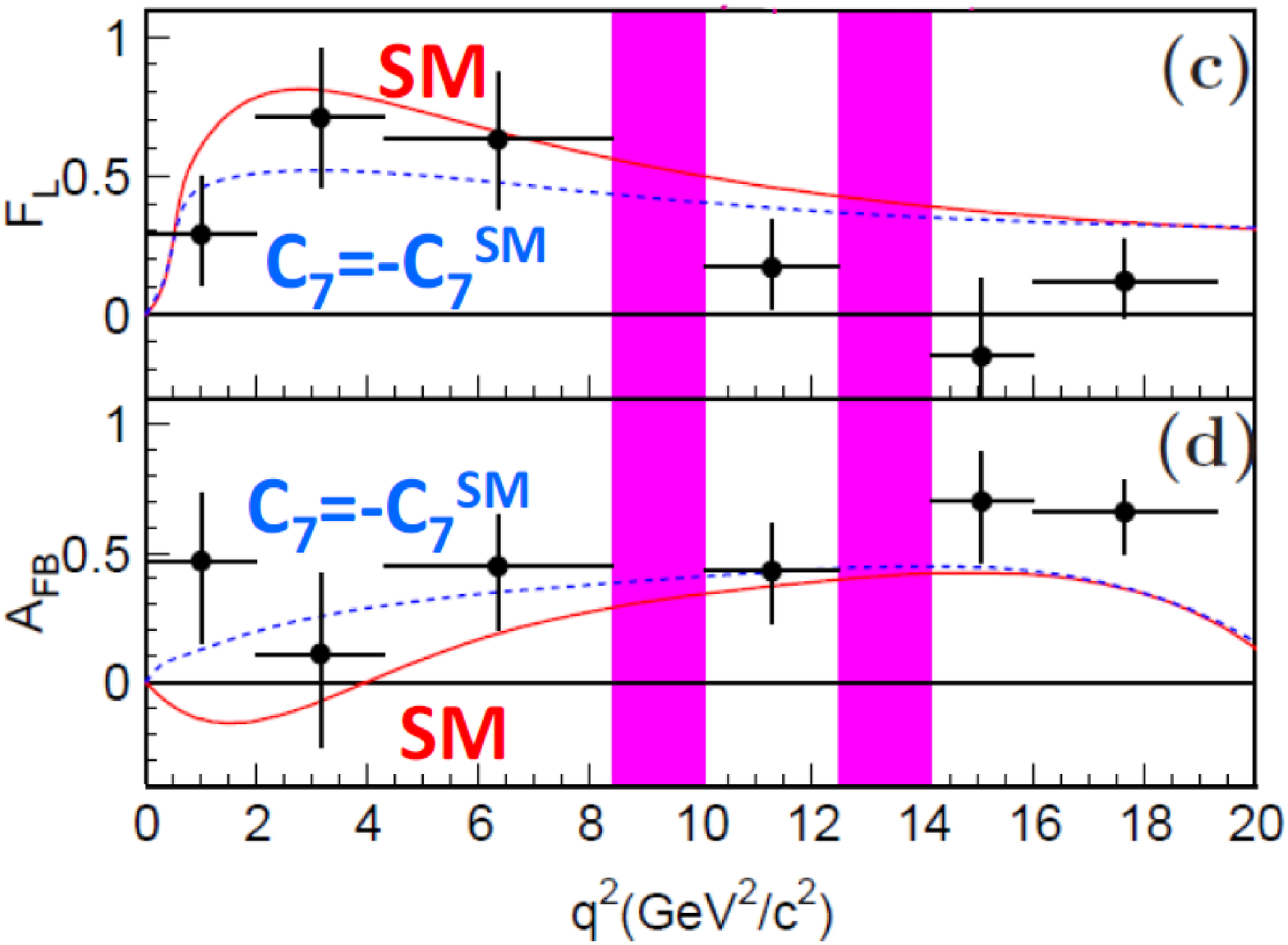}
   \includegraphics[width=0.35\textwidth]{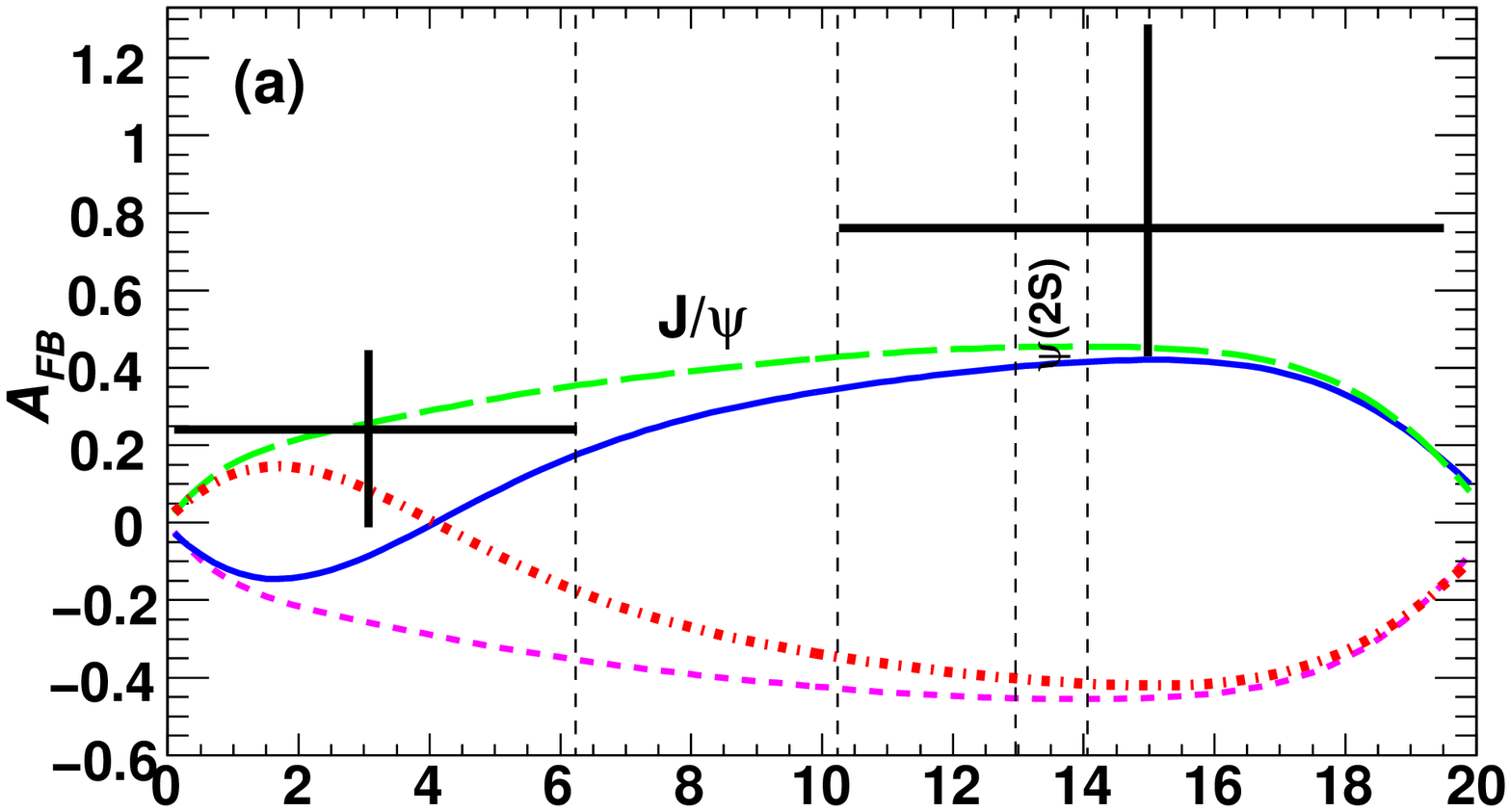}}
\caption{The forward-backward asymmetry of the muon in
$B\rightarrow K^*\mu+\mu^-$ decays as a function of dimuon
invariant mass from CDF (left), Belle (middle) and \niceMKbabar{}
(right). The points represent measurement, the red line in CDF
and Belle case and the blue line in \niceMKbabar{} result show
the standard model prediction and the other curves represent
a different beyond standard model scenarios. The areas without
data points correspond to the charmonium regions which are
excluded from the analysis.}\label{fig:afb}
\end{figure}

Another example of an FCNC rare process which generates lot of
excitement these days is a class of the decays governed by the
$b\rightarrow sl^+l^-$ quark level transition with $l$ being
a charged lepton. Decays $B^{0,\pm}\rightarrow
K^{0,\pm}\mu^+\mu^-$ and $B^{0,\pm}\rightarrow
K^{*0,\pm}\mu^+\mu^-$ were already observed. Recently CDF
experiment observed also decay
$B_s\rightarrow\phi\mu^+\mu^-$ with $\approx 6.3\sigma$
significance using 4.4 fb$^{-1}$ of data \cite{cdf:bkmumu}. 
The measured branching fraction is
$(1.44\pm0.33\pm0.46)\cdot10^{-6}$. As those decays proceed
even in the standard model through more than one amplitude,
there is a rich phenomenology of interferences. From the
interference effects, the forward-backward asymmetry of the
muons as a function of dimuon invariant mass is the one which is
responsible for the excitement. It is measurement in Belle
\cite{Wei:2009zv}, \niceMKbabar{} \cite{Aubert:2008ju} and CDF \cite{cdf:bkmumu} experiments and we
show results in Fig.~\ref{fig:afb}.
While not statistically significant, all three experiments
show some departure in the same direction from the standard model.
It is going to be interesting to follow future measurements of this
quantity.

\section{Conclusions}

\begin{figure}
\centerline{\includegraphics[width=0.3\textwidth]{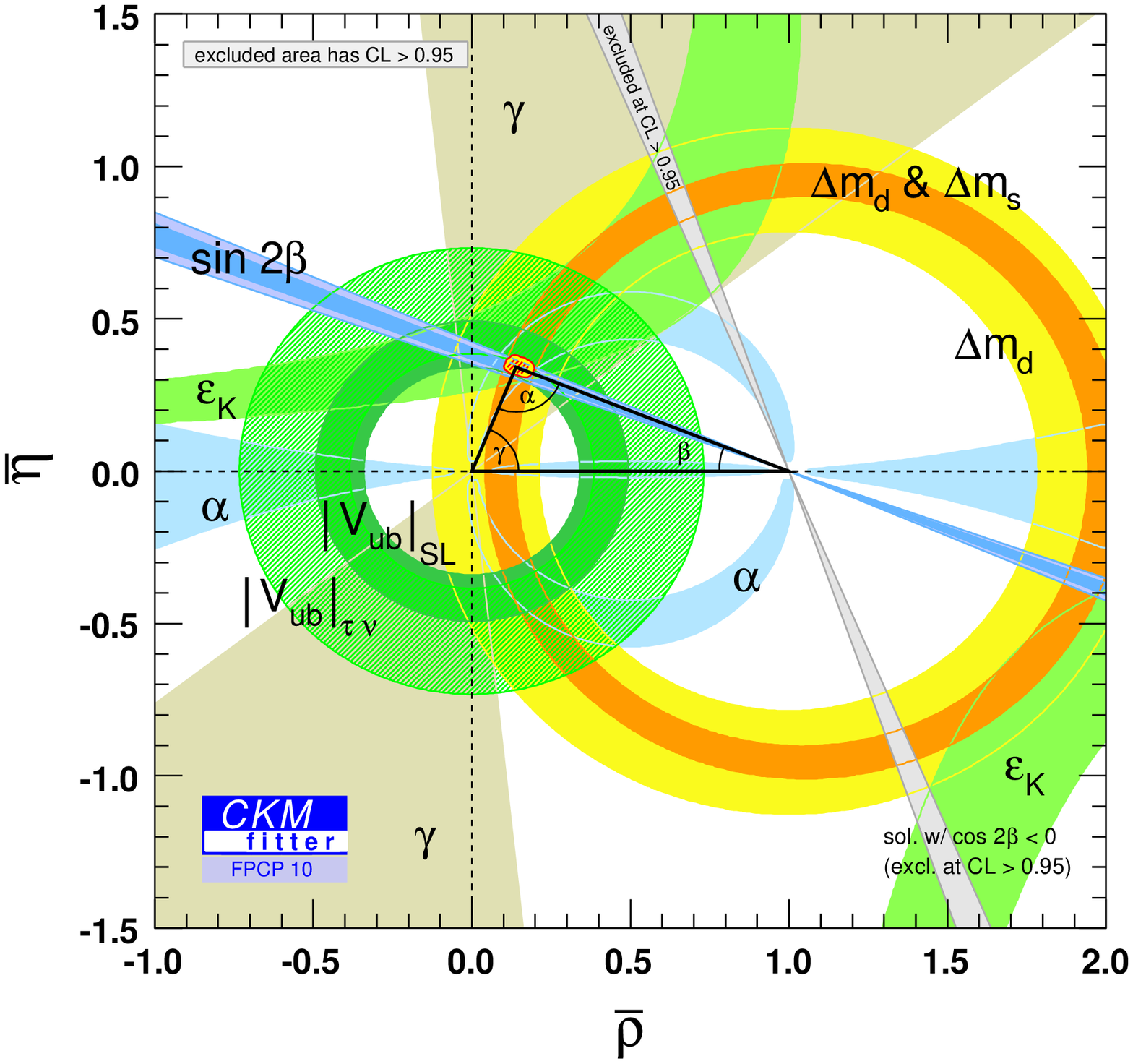}
  \includegraphics[width=0.30\textwidth]{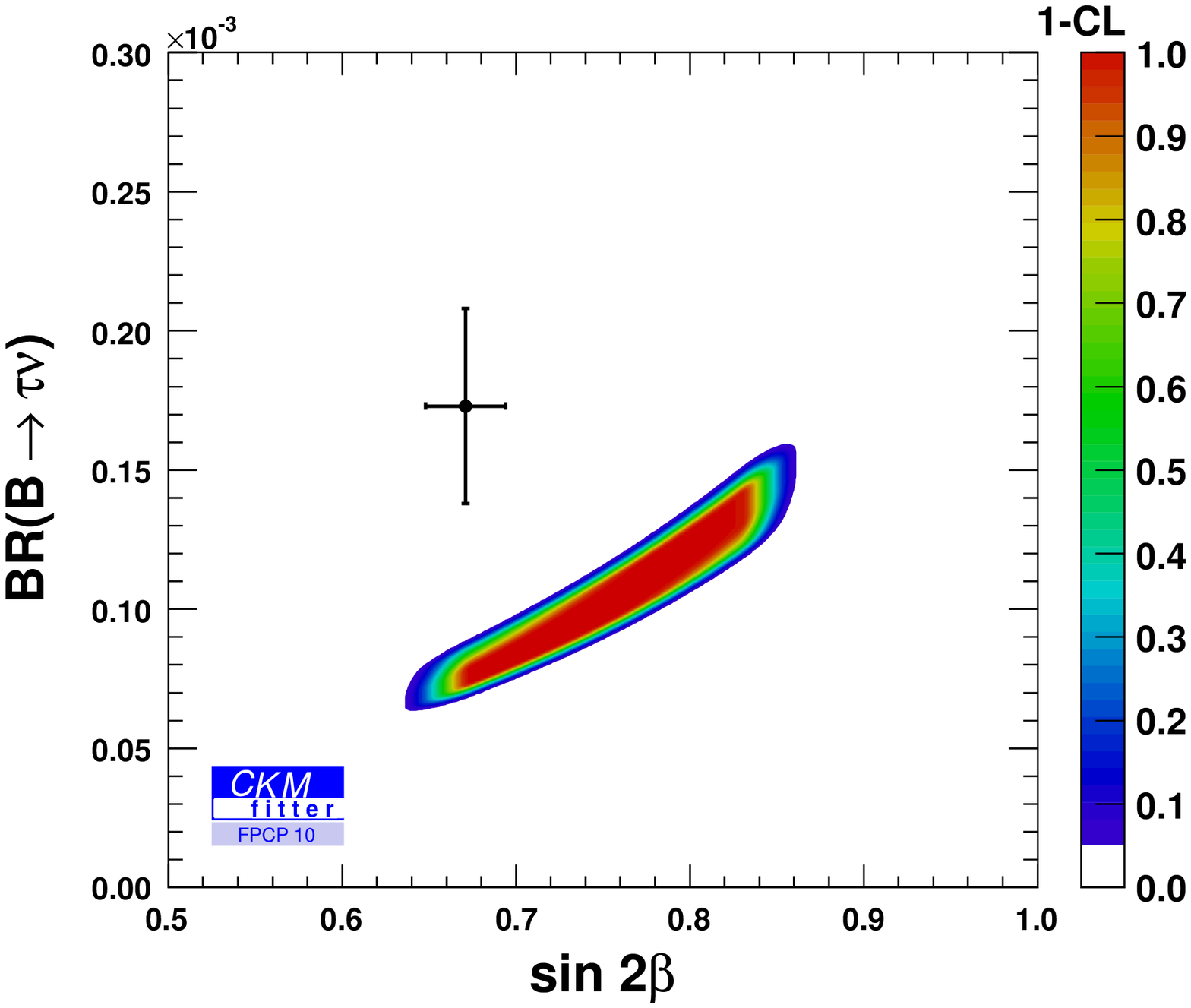}
   \includegraphics[width=0.30\textwidth]{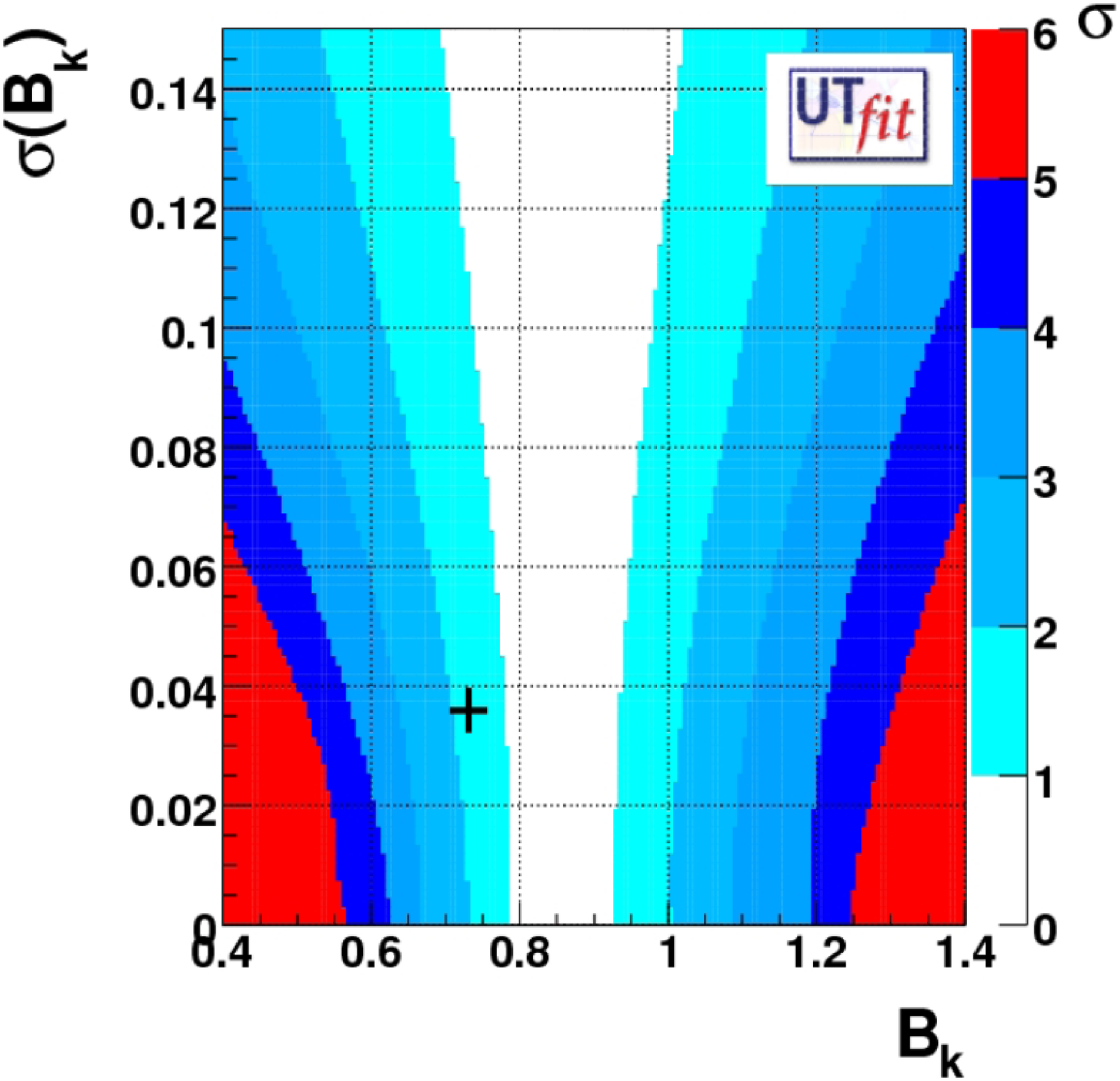}}
\caption{The global status of the unitarity triangle fit from
CKMFitter group (left), the graphical representation of the
$B\rightarrow \tau\nu$ versus $\sin2\beta$ disagreement
(middle) and the situation with indirect and direct
determinations of the parameter proportional to $\epsilon_K$
from UTFit group (right). In the middle plot, colored
confidence regions show expectation for the $B\rightarrow
\tau\nu$ branching fraction from the fit where two quantities
are excluded while the point shows experimental results. In the
right plot, the colored areas show confidence region of the $B_K$
from the fit without constraint from the $CP$ violation in the kaon
system and the cross represents experimental measurement of the
quantity.}\label{fig:ckmfit}
\end{figure}

Globally, except of the flavor specific asymmetry in
semileptonic $b$-decays, there is not a significant
discrepancy in the global picture of $CP$ violation. On the
other hand, there are few discrepancies which are worth to
follow in the future. In Fig.~\ref{fig:ckmfit} we show
the global status of the CKM fit \cite{Charles:2004jd}. Other
determination \cite{Ciuchini:2000de,Lunghi:2009ke} provide similar picture.
All groups see $\approx 2.5\sigma$ improvement of the fit
if either constraint from the $B\rightarrow \tau\nu$ or $\sin2\beta$ is
removed from the fit. Other main small discrepancies are in
the $V_{ub}$ and the $CP$ violation parameter $\epsilon_K$
in the kaon system. It is worth to note that the discrepancy
between measured $\sin(2\beta)$ and its prediction from the
fit without $\sin(2\beta)$ was pointed out already in 2007
\cite{Lunghi:2007ak}.

On the limited space we could not discuss the charm quark
sector, which has strong potential. Its status and prospects
at the time of conference can be find in
Ref.~\cite{proc:Mannel}. The prospects of the LHC in the bottom quark
sector were discussed in several contribution, with most
relevant one with respect to this work being
Ref.~\cite{proc:Conti}. With large expectations whole
community is positive about future interesting results and
the importance of the flavor physics for discovering and/or
understanding a physics beyond standard model.

\section*{Acknowledgments}

The author would like to thank organizers for the kind
invitation to the conference and would like to acknowledge
support from the Bundesministerium f\"ur Bildung und
Forschung, Germany. 


\begin{footnotesize}

\end{footnotesize}


\end{document}